IAC-17-D5.4.4

# SPOOQYSATS: CUBESATS TO DEMONSTRATE QUANTUM KEY DISTRIBUTION TECHNOLOGIES

James A. Grieve[a], Robert Bedington[a]*, Rakhitha C.M.R.B. Chandrasekara[a], Alexander Ling[a,b]

[a] *Centre for Quantum Technologies, National University of Singapore, Block S 15, 3 Science Drive 2, Singapore, 117543*, r.bedington@nus.edu.sg
[b] *Department of Physics, 2 Science Drive 3 Blk S12, Level 2, Singapore 117551, Singapore*,
* Corresponding Author

**Abstract**
Satellite-based QKD offers the potential to share highly secure encryption keys between optical ground stations all over the planet. SpooQySats is a programme for establishing the space worthiness of highly-miniaturised, polarization entangled, photon pair sources using CubeSat nanosatellites. The sources are being developed iteratively with an early version in orbit already and improved versions soon to be launched. Once fully developed, the photon pair sources can be deployed on more advanced satellites that are equipped with optical links. These can allow for very secure uplinks and downlinks and can be used to establish a global space-based quantum key distribution network. This would enable highly secure symmetric encryption keys to be shared between optical ground stations all over the planet.

## 1. Acronyms/Abbreviations
Quantum Key Distribution (QKD)
Advanced Encryption Standard (AES)
Technology Readiness Level (TRL)
Size, Weight and Power (SWaP)
Small Photon Entangling Quantum System (SPEQS)

## 2. Introduction
The ongoing development of quantum computers threatens traditional cryptographic architectures, with public key cryptosystems (for example, the RSA algorithm [1]) particularly vulnerable [2]. Symmetric architectures such as the Advanced Encryption Standard (AES) [3] may be able to preserve security in a post-quantum world at the cost of increased key sizes and key refresh rates, placing new emphasis on key management and distribution systems. Quantum Key Distribution (QKD) seeks to address this need via a family of techniques based on the transmission of weak quantum signals, in which symmetric keys can be established between two or more parties in a manner resistant to any computational attack [4,5].

All QKD techniques employ at least two channels for communication, one of which is used to transmit quantum signals which may be either extremely weak optical signals at the level of single photons [4], or single photons linked by the property of quantum entanglement [5]. This "quantum channel" is particularly adversely affected by channel loss, severely restricting the distance over which such techniques can be deployed terrestrially. In optical fibers typical transmission losses restrict QKD demonstrations to at most a few 100s of kilometers (using exotic low-loss optical fibers [6]). Over long enough distances, free-space transmission yields favourable results, though ultimately this is limited by line-of-sight constraints (the terrestrial record of 144km [7] took place between island mountain summits). Satellite-based QKD schemes offer an elegant solution to these problems, and provide a clear pathway towards the establishment of global-scale key distribution networks.

Recently, high profile QKD experiments using satellites have been reported [8,9] employing both "first generation" schemes (e.g. the BB84 protocol [4]) alongside the distribution of entangled photons between widely separated ground stations [8] – a prerequisite for second-generation QKD protocols. These experiments have attracted considerable attention, and demonstrate unambiguously the feasibility of satellite-based quantum key distribution.

At the National University of Singapore's Centre for Quantum Technologies, we have pursued an iterative approach to this problem, developing lightweight and rugged entangled photon pair sources with sufficiently low SWaP requirements to allow hosting on board "CubeSat" class nanosatellites. As their popularity increases, nanosatellites and the CubeSat platform continue to represent an attractive pathway towards establishing flight heritage and raising the technology readiness level (TRL) of components in our QKD system. The SpooQySat programme consists of a number of field tests and satellite missions tied to the parallel development of a miniature, robust and highly capable entangled photon source known as SPEQS [10].

In this paper, we briefly review proposed satellite QKD architectures and the progress being made at CQT developing and testing our quantum light sources.

### 3. Satellite QKD Schemes

To share quantum encryption keys between space and ground the quantum light source can be located on the ground or in space (see Figure 1 and [12]), though for maximum utility the source should be located in orbit (three out of four options in Figure 1 require this). Beaming photons from space-to-ground also has the advantage that the optical transmission path is much less affected by atmospheric turbulence (because the effects of turbulence act primarily at the end of the transmission, where the atmosphere is thickest) so optical link losses are lower [13]. Despite this, the perceived difficulty in operating a quantum source on-board a satellite has led several groups to propose ground-to-space missions [14 - 17].

If scenarios 1 or 2 are used to share keys between two ground stations then it is required that the satellites be trusted parties in the key exchange (we call these "trusted nodes"). In contrast, scenario 3 (recently demonstrated in the Micius mission [8]) potentially enables entanglement-based QKD, removing this trust requirement. This is because if one photon in each pair of photons is beamed to Alice (at one ground station) and the other of the pair beamed to Bob (who is at the other ground station) then only Alice and Bob will have knowledge of the measurement outcomes and the key generated is private to them. They will be able to test this quantum exclusivity by performing a Bell violation test [5], and consequently need not trust the source. This scheme is restricted to ground stations simultaneously within the field-of-view of the satellite, and since both photons in a pair are traveling along high-loss paths the probability of both photons in a pair reaching their ground stations is very small (~60dB of total loss). Consequently key generation rates would be very slow, and indeed the recent demonstration of entanglement distribution by the Micius mission was able to achieve only slightly more than a single pair per second [8].

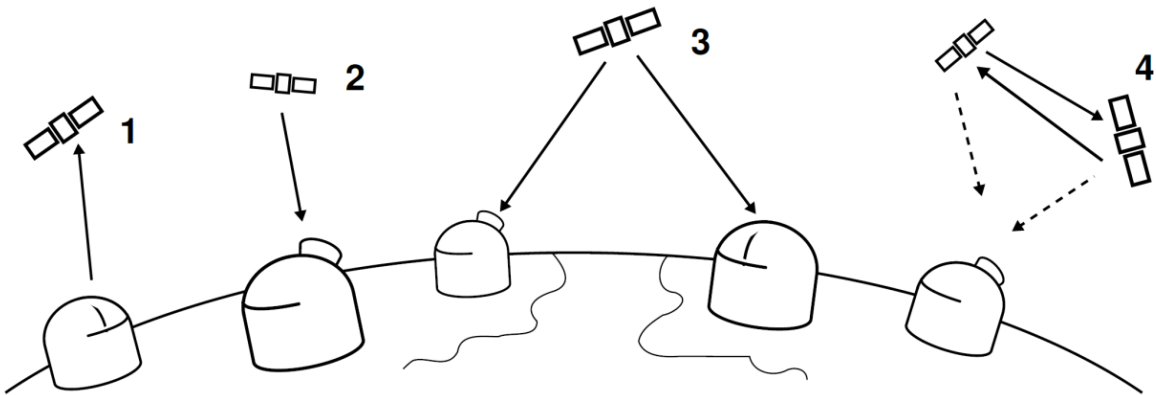

Figure 1- Possible satellite-based QKD implementations. 1. Ground-to-space, where the photon source is on the ground and the satellite only carries detectors. 2. Space-to-ground, where the satellite carries a source and detectors. 3. A platform that can beam down to two ground stations simultaneously. 4. Inter-satellite QKD which could be the building block for a long baseline test of quantum correlations. To enable configurations 2-4 with Bell violation-type measurements, a source of entangled photons in space must be demonstrated. Figure reproduced from [11].

### 4. Trusted node configuration

As detailed in the previous section, practical key generation schemes using satellites in the near term will likely operate in a trusted node configuration (see Figure 1). In this scenario, the satellite carries out QKD operations with ground stations on an individual basis to establish independent secret keys with each of them. The satellite holds all keys, while the stations only have access to their own keys. To enable any pair of stations to share a common key, the satellite combines their respective keys $K_A$ and $K_B$ and broadcasts their bit-wise parity $K_A \oplus K_B$. Since the original keys are independent secret strings, their bit-wise parity is a uniformly random string, and knowledge of the parity does not reveal any useful information to an adversary. Using this information along with knowledge of their own secret key, the stations can extract the opposite key via $K_A \oplus (K_A \oplus K_B) = K_B$ and $K_B \oplus (K_A \oplus K_B) = K_A$. Of course, since the satellite can access all keys, access to the satellite would give an adversary complete knowledge of any key and hence the satellite must be trusted.

### 5. Developments at the Centre for Quantum Technologies, National University of Singapore

While the optical links and optical ground stations required for satellite QKD can be adapted from existing lasercomms systems, space-based quantum key sources require more development work. The most secure forms of QKD require entangled photon pairs, and CQT are developing SPEQS (the Small Photon Entangling Quantum System) to address the need for a robust, space qualified source of entanglement. SPEQS is being developed in iterative cycles that include regular on-orbit testing in CubeSats, as outlined in Table 1.

The aim is to gradually raise the technology readiness level of the photon pair sources. Initial iterations developed the miniaturized electronics and ruggedized optics. These led onto on-orbit tests of a correlated photon pair source [22] (these photons are not entangled in polarization but may have entanglement in other degrees of freedom, e.g. time-frequency). A polarization entangled version of SPEQS is being developed for launch in the SpooQy-1 CubeSat in 2018 [10]. These CubeSat tests will establish the space worthiness of the source, but they do not beam photons out of the satellite to perform QKD. Once space worthiness is established, the first missions will target cases 2 and 4 from Figure 1. To this end, we have joined with a consortium of researchers and industry partners to put forward a proposal for a satellite-to-ground entanglement distribution and QKD mission (CQuCoM, [23]). This ambitious programme will make use of CubeSats as the orbiting platform, and if successful would provide invaluable insight into the minimal SWaP required for such protocols. We have also established a partnership with researchers at the University of New South Wales to produce a case study for a mission to explore satellite-to-satellite entanglement distribution using CubeSats [24]. While not yet demonstrated, we anticipate that inter-satellite links of this sort will be critical components of future QKD systems [25], enabling a constellation of satellites operating under the "trusted node" paradigm to dramatically increase the rate at which distant ground stations can establish key pairs.

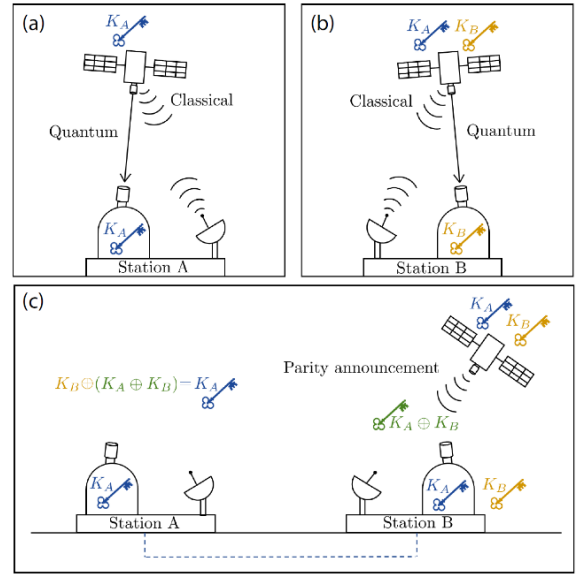

Figure 2- Illustration of the most common satellite QKD scheme: the downlink trusted-node. (a) A QKD system on-board a satellite performs a key exchange operation during its pass over a ground station (Station A), establishing a secret key pair $K_A$. (b) The satellite repeats this procedure during overflight of a second ground receiver (Station B) establishing the key pair $K_B$. (c) The satellite broadcasts the bitwise parity of KA and KB over a classical channel. This reveals no information to an adversary while enabling both Station A and Station B to extract the opposite key pair. Figure reproduced from [18].

Table 1: Milestones leading to on-orbit QKD demonstrations.

| Year | Milestone | Mission | Reference |
|---|---|---|---|
| 2012 | Basic miniature SPDC source. | High altitude balloon Test | [19] |
| 2013 | Correlated SPDC pair source. | High altitude balloon Test | [20] |
| 2014 | Space-qualified, correlated, SPDC source survives rocket explosion | GomSpace GomX-2 CubeSat | [21] |
| 2016 | Space-qualified, correlated, SPDC source in low Earth Orbit | NUS Galassia CubeSat | [22] |
| 2018 | Demonstration of entangled photon pair source in space. | CQT SpooQy-1 CubeSat | [11] |
| *2019+* | *QKD demonstrations with optical links in space.* | *International collaborations* | *[23]* |

## 6. Conclusions

Quantum key distribution offers a technological solution to the problem of ensuring privacy in a post-quantum world, and seems likely to play a role in future cryptosystems. As QKD schemes are all reliant upon the transmission of quantum signals, channel loss represents a crucial barrier to its adoption for long-range communications, motivating many researchers to propose satellite-based architectures. As recent technology demonstrations [8] have firmly established the technological feasibility of this approach, we continue to follow an iterative development scheme, working towards small and cost-effective QKD terminals.

The SpooQySat programme described in this paper encompasses a series of high altitude and on-orbit tests of the miniaturized photon pair source SPEQS, and aims to produce a fully validated system for the production of entanglement in the satellite environment. Once this is accomplished, the source will be well positioned for incorporation into fully fledged QKD systems, either on dedicated satellite missions (implementing, for example the trusted node architecture described in Figure 2) or as QKD terminals to increase the capability of larger multifunctional platforms. We are actively developing international partnerships to explore opportunities in this area, and hope to participate in the first QKD missions very soon.

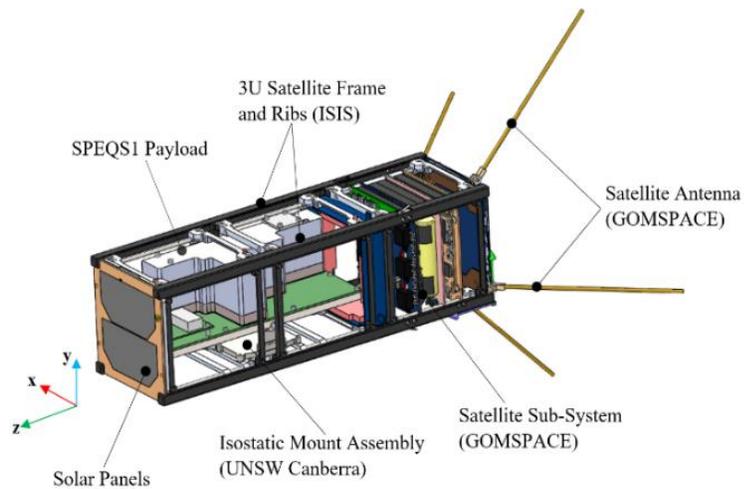

Figure 3 Rendering of the SpooQy-1 spacecraft (with most solar panels and cables removed).
Approximately 2/3 of the volume is occupied by the SPEQS payload.

**Acknowledgements**

This work is partially supported by the National Research Foundation, Prime Minister's Office, Singapore (under the Research Centres of Excellence programme and through Award No. NRF-CRP12-2013-02) and by the Singapore Ministry of Education (partly through the Academic Research Fund Tier 3 MOE2012-T3-1-009).